\documentclass[prl,aps,twocolumn,showpacs,superscriptaddress,floatfix]{revtex4}
\usepackage{graphicx}
\usepackage{bm}

\usepackage[dvips]{color}
\usepackage{graphicx}

\begin{document}

\title{Accurate determination of the Gaussian transition in spin-1 chains
with single-ion anisotropy}

\author{Shijie Hu}
\affiliation{Department of Physics, Renmin University of China, Beijing
100872, China}
\affiliation{Institut f\"ur Theoretische Physik, Georg-August-Universit\"at
Goettingen, 37077 G\"ottingen, Germany}
\author{B. Normand}
\affiliation{Department of Physics, Renmin University of China, Beijing
100872, China}
\author{Xiaoqun Wang}
\affiliation{Department of Physics, Renmin University of China, Beijing
100872, China}
\author{Lu Yu}
\affiliation{Institute of Physics, Chinese Academy of Sciences, Beijing
100190, China}

\date{\today}
\begin{abstract}
The Gaussian transition in the spin-one Heisenberg chain with single-ion 
anisotropy is extremely difficult to treat, both analytically and numerically. 
We introduce an improved DMRG procedure with strict error control, which we 
use to access very large systems. By considering the bulk entropy, we 
determine the Gaussian transition point to 4-digit accuracy, $D_{c}/J
 = 0.96845(8)$, resolving a long-standing debate in quantum magnetism. 
With this value, we obtain high-precision data for the critical behavior 
of quantities including the ground-state energy, gap, and transverse 
string-order parameter, and for the critical exponent, $\nu = 1.472(2)$. 
Applying our improved technique at $J_{z} = 0.5$ highlights essential 
differences in critical behavior along the Gaussian transition line.

\end{abstract}



\maketitle


The Gaussian transition appears in several fields of quantum physics and
statistical mechanics. The equivalence between surface-roughening transitions
in classical two-dimensional (2D) models and quantum phase transitions in
spin chains was introduced in Ref.~\cite{rls}, and their rich phase diagrams
investigated at length in Ref.~\cite{rdnr}. Characterized by continuously
variable exponents, the Gaussian transition differs significantly both from
regular phase transitions and from those of Kosterlitz-Thouless (KT) type.
These differences complicate both analytical and numerical approaches to a 
complete and accurate description of rough surfaces and quantum spin chains.

The $S = 1$ Heisenberg chain is one of the fundamental models in quantum
magnetism. It formed the basis of Haldane's conjecture \cite{haldane} for
a finite gap in antiferromagnetic chains with integer spin, as
opposed to the gapless spectrum of half-odd-integer cases. Numerically,
quantum spin chains are important test-cases for any computational
technique, and Haldane's prediction has been verified by a range of methods
with increasing accuracy \cite{blote,white}. Experimentally, while the
``Haldane gap'' has been found in the excitation spectra of many systems
\cite{rybno}, most known $S = 1$ chains, including NENP \cite{rnenp}, 
NINAZ \cite{rninaz}, and NDMAP \cite{rndmap}, are organic Ni materials with 
significant single-ion anisotropies. Analytical approaches to the Gaussian 
transition driven by this term are complicated by the lack of a suitable 
effective field theory \cite{rs}, and its broad nature makes all numerical 
techniques difficult to apply. Many authors have considered this transition, 
producing occasionally contradictory results \cite{rbjk,rst,rgjl,rtnk,rchs,
rdeor,rthcy,runk,raho}.

In this Letter we resolve the problem of the Gaussian transition in the 
$S$ = 1 chain with single-ion anisotropy. We exploit the fact that this 
transition is a gapless point between two gapped phases, whence the 
entropy exhibits a sharp peak. We introduce an improved density-matrix 
renormalization-group (DMRG) approach with systematic error control, 
allowing high-precision calculations at system sizes up to 
$L = 20000$, which automatically eliminate the end-spin entropy. We
determine the critical point with very high accuracy, and thereby deduce
the critical behavior of several quantities at different points on the
Gaussian transition line.

The general form of the model is
\begin{equation}
{\cal H} \! = \! \mbox{$\sum^{L}_{i=1}$} \! J (S^{x}_{i} S^{x}_{i+1} \! + \! 
S^{y}_{i} S^{y}_{i+1}) + J_{z} S^{z}_{i} S^{z}_{i+1} + D \left( S^{z}_{i} 
\right)^{2} \!\!
\end{equation}
where $J_z$ interpolates between XY and Ising spins, $D$ is the single-ion
anisotropy, and $L$ the length of the chain. The full parameter space of
$(D,J_z)$ contains N\'eel, Haldane, large-$D$, ferromagnetic, and two XY
phases. In classical planar surface, or ``solid-on-solid,'' models, the 
N\'eel and large-$D$ phases are different ``flat'' phases, the Haldane 
phase is ``rough,'' and the Gaussian transition is of ``preroughening'' 
type. These are the three phases of the $S = 1$ Heisenberg chain ($J_z
 = 1$) as $D$ is varied. While the N\'eel phase possesses $Z_2$ symmetry 
and the Haldane phase an incomplete $Z_2 \times Z_2$ symmetry, the large-$D$ 
phase has no remaining symmetries. The Gaussian transition is a line in the $
(D,J_z)$ plane, on which the excitations are gapless. This line is well 
described by a conformal field theory (CFT) \cite{cft}, and has been analyzed 
in a number of studies \cite{rchs,rdeor,rthcy,runk}, but none has achieved 
the numerical precision required for a consistent discussion of the critical 
behavior across the transition.

DMRG is the most efficient and accurate numerical technique for 1D systems 
\cite{white}. Anticipating the need for both large system sizes and extreme 
precision, we begin by introducing an improved DMRG technique. In the 
conventional scheme, the absolute (coupled round-off and truncation) 
error increases systematically with $L$, and this accumulated error has a 
strong effect on the relability of the computation, possibly even 
disguising the critical behavior in a quantum many-body system. We fix the 
round-off error by renormalizing the lowest eigenvalue of Hamiltonian to 
remain of order 1, thereby obtaining a very significant reduction in the 
truncation error for large systems.
In the DMRG iteration, we replace the original Hamiltonian matrix $H(m,L)$, 
for chains of $L$ sites with $m$ kept states, by $H(m,L) - [\varepsilon_1 
(m, L-2) - \delta]$, where $\varepsilon_1 (m, L-2)$ is the lowest eigenvalue 
of $H (m,L-2)$ and $\delta$ is a constant chosen such that $\varepsilon_1 
(m,L) \sim O(1)$. Here we use $\delta/J = 1$ throughout. While the (extensive) 
total energy of the ground state, $E_g (m,L)$, can be reconstructed by 
summation, its (intensive) average value per site is determined directly and 
self-consistently as $e_g (m,L) = [\varepsilon_1 (m,L) - \delta]/2$. 
Similarly, for the first excited state $e_f (m,L) = [\varepsilon_2 (m,L) - 
\epsilon_2 (m,L-2)]/2 + e_g (m,L)$, where $\varepsilon_2 (m,L)$ is the 
second-lowest eigenvalue of $H (m,L)$.

The gap in our method is given simply by $\Delta (m,L) = \varepsilon_2 (m,L) 
 - \varepsilon_1 (m,L)$. Its general expression is   
\begin{equation}
\Delta (m, \! L) \! = \! [e_f ( \! m \! ) - e_g ( \! m \! )] L + \Delta 
( \! m \! ) + \! \mbox{$\sum^{\infty}_{n = 1}$} \! \frac{\alpha_n ( \! m 
\! )}{L^n}, 
\label{egl}
\end{equation}
where $e_g (m)$ and $e_f (m)$ are the intensive energies of the ground and 
first excited states for infinite $L$, and become equal for infinite $m$. 
In the polynomial expansion of contributions at higher order in $1/L$, 
the $n = 1$ term arises from truncation errors and open boundary conditions 
(OBCs), while the $n = 2$ term has contributions from fluctuations at the 
quadratic band minimum. Here we calculate the energies in the linear term 
independently by extrapolation. Subtracting these gives a gap function 
$\Delta (m,L)$ that decreases monotonically with increasing $L$. A second 
polynomial fit of $\Delta (m)$ allows its extrapolation to infinite $m$ to 
obtain the true gap. 

Sharing its foundations with quantum information theory, the DMRG method 
is ideally suited to discussions of entropy and entanglement. The von 
Neumann entropy, ${\cal S} (m,L) = - {\rm Tr} \rho (m,L) \ln \rho (m,L)$, 
is readily computed from the reduced density matrix, which we obtain to 
high accuracy throughout our calculations with the renormalized Hamiltonian. 
The entropy obeys an area law except in critical regimes, where it depends 
logarithmically on $L$ \cite{rcc}. This extremum in entropy is an excellent 
indicator of a (gapless) critical point between two gapped phases.

Before analyzing the entropy, we discuss the special and remarkable feature 
of the $S = 1$ Heisenberg chain, that free $S = 1/2$ entities are found at 
a chain end, both in theory and in experiment \cite{effectivespin}. In the 
Haldane phase with OBCs, the two free end-spins can be described by 
$\hat H_{\rm eff} = J_{\rm eff} {\vec S}_L \cdot {\vec S}_R$ \cite{white}, 
where the effective coupling $J_{\rm eff} > 0$ falls exponentially with 
$L$. In the Hilbert space $S^{\rm tot}_z = 0$, the two spins are maximally 
entangled with entropy $\ln 2$, while for $S^{\rm tot}_z = 1$ they are 
unentangled. The additional truncation error due to this edge-entropy 
contribution causes significant computational difficulties. In Fig.~1(a) 
we find a $\ln 2$ drop in the ground-state entropy ${\cal S} (L)$ in the 
Hilbert space $S^{\rm tot}_z = 0$ when the chain reaches a certain length 
at fixed $D$. For $D = 0.92$ and $m = 1000$, this occurs at $L = 4500$ 
[inset, Fig.~1(a)]. When $L$ becomes sufficiently large, $J_{\rm eff}$ 
falls below the machine precision and the end-spin contribution vanishes. 
The remaining ``bulk'' entropy contains the essential physics of the spin 
chain. Different but conceptually similar approaches have considered both 
the two-site entropy and ${\cal S} (L)$ in a chain with no end-spin effects 
\cite{rlstn}. 

\begin{figure}[t]
\includegraphics[width=8.5cm]{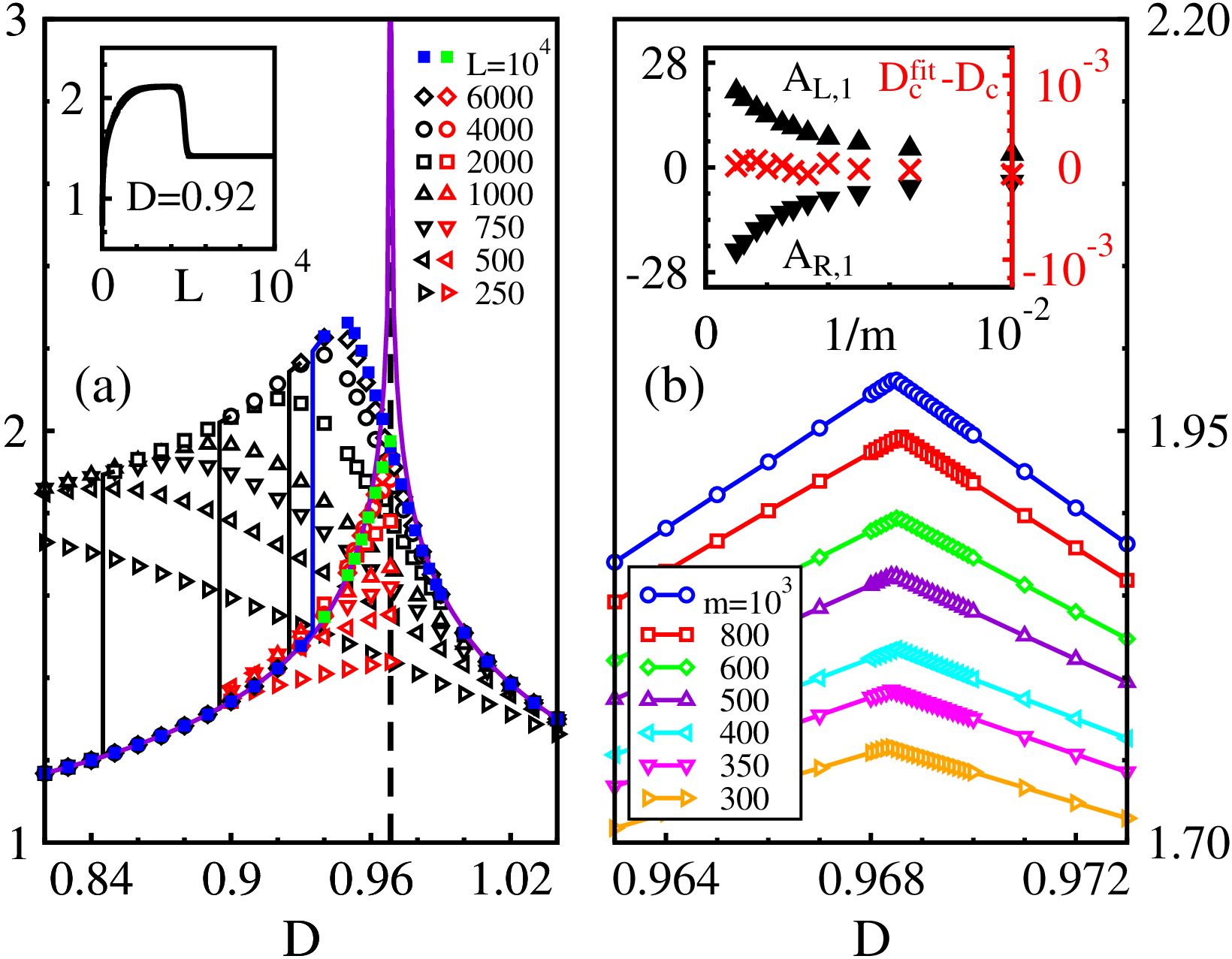}
\caption{(color online) Entropy ${\cal S}$ as a function of $D$ for $J_z
 = 1$. (a) Calculations with $m = 1000$. Open symbols are obtained for 
the lowest energy level in Hilbert space $S^{\rm tot}_z = 0$ with a range 
of $L$ values, solid symbols for $S^{\rm tot}_z = 1$. Inset: $\ln 2$ drop 
in ${\cal S}(L)$ for $D = 0.92$. (b) Bulk entropy ${\cal S}(D)$ close to 
the Gaussian critical point, computed with $L = 10000$ for a range of $m$ 
values. Insets: fitting slopes $A_{L,1}$ and $A_{R,1}$ (left axis) and 
transition $D_{c}^{\rm fit}$ (right) obtained as functions of $m$.}
\label{fig1}
\end{figure}

Figure 1(a) contrasts the total and the bulk entropy. Calculations with small 
$L$ cannot access the unentangled regime, and for larger $L$ we find a $\ln 
2$ jump when $D$ approaches $D_c$. For $L = 10000$, the end-spins remain 
entangled for $0.94 < D < D_c$. The maximum in the total entropy moves 
strongly with $L$, showing no direct indication of criticality \cite{rthcy}. 
By contrast, in the Hilbert space $S^{\rm tot}_z = 1$, the end-spins are 
unentangled in the lowest-energy state and this data reproduces exactly the 
bulk entropy. The location of the maximum in ${\cal S}$, shown in detail in 
Fig.~1(b), is clearly invariant with $m$. A linear fit to the bulk entropy 
on both sides of the transition in Fig.~1(b) gives our primary result, 
$D_{c}/J = 0.96845$ with a minuscule error bar of 0.00008. The increasing 
slopes of the bulk entropy lines as $m \rightarrow \infty$ [inset Fig.~1(b)] 
indicate the onset of critical behavior.

\begin{figure}[t]
\includegraphics[width=8.5cm]{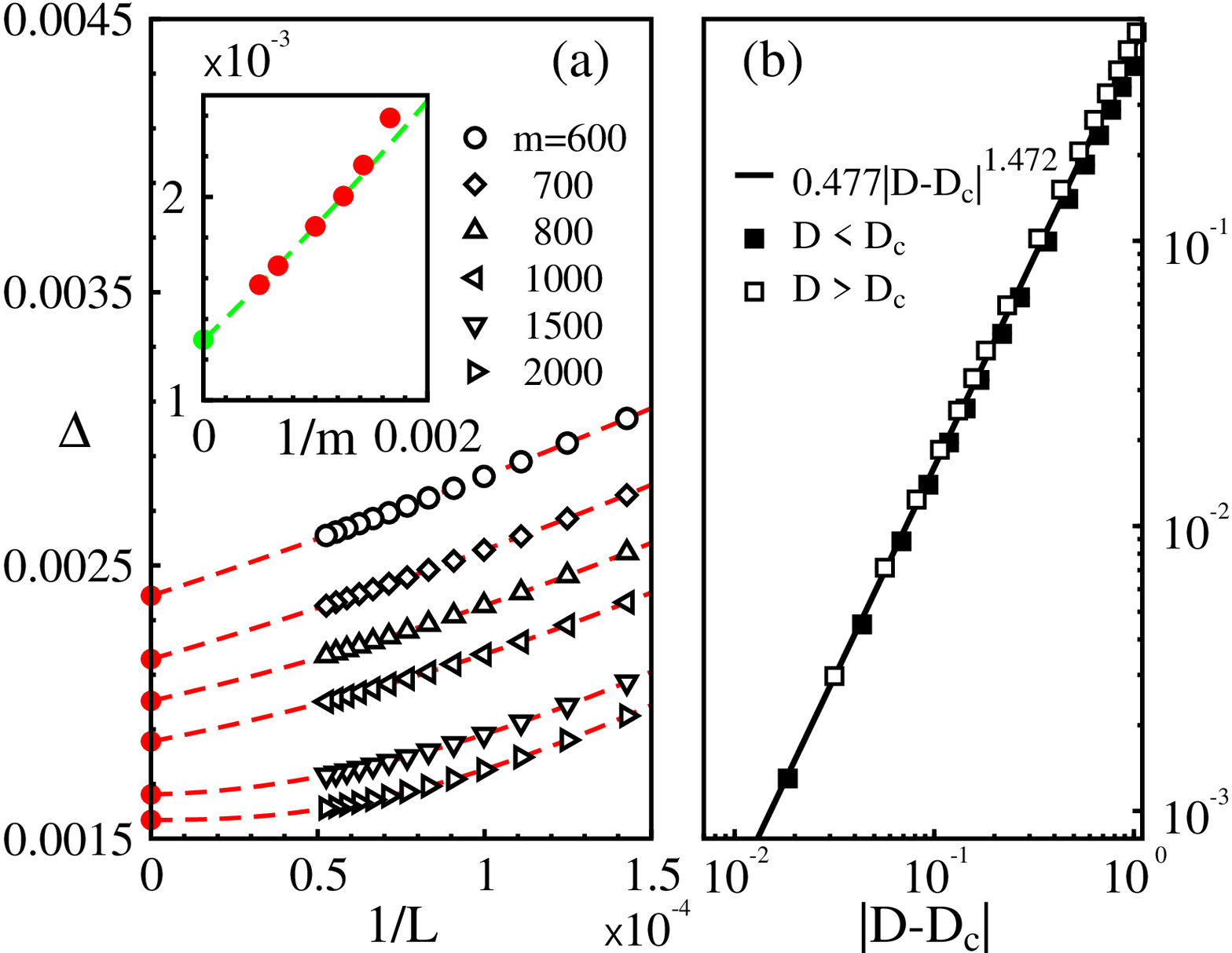}
\caption{(color online) (a) Gap as a function of $L$, computed for $D/J 
 = 0.95$ with several values of $m$. Solid symbols for $L \rightarrow 
\infty$ are extrapolated to $m \rightarrow \infty$ (inset), giving 
$\Delta (0.95) = 0.00130(4)$. (b) Extrapolated gaps as a function of 
$\left| D - D_{c} \right|$.}
\label{fig2}
\end{figure}

Having determined this extremely precise value of $D_c$, we may now discuss 
the critical behavior of the Gaussian transition with hitherto unattainable 
accuracy. We consider the physical quantities used in previous analyses of 
the transition \cite{rbjk,rst,rgjl,rtnk,rchs,rdeor,rthcy,runk,raho}, beginning 
with the gap. To avoid effects in the gap extrapolation related to the 
disappearance of edge states, we use the lowest energy levels in the Hilbert 
spaces $S^{\rm tot}_z = 1$ and $S_z^{\rm tot} = 2$. Figure 2(a) illustrates 
our two-step extrapolation approach to compute the gap for the extremely 
numerically challenging point $D = 0.95$, which lies very close to $D_c$. 
By following this procedure for all values of $D$, we show 
in Fig.~2(b) the approach of the gap to zero at $D_c$ from both the Haldane 
and large-$D$ sides. The closest four points, $D = 0.925$, 0.95, 1.0, and 
1.025, reveal a very narrow critical region, $|D - D_c| < 0.1$, with critical 
exponent $\nu = 1.472(4)$.

In a CFT for the Gaussian critical line \cite{cft}, the gap $\Delta$
varies linearly and the energy $e_{g}$ quadratically with $1/L$. For the 
CFT analysis, we perform DMRG calculations with periodic BCs (PBCs) using 
$L = 200$ and $m = 2000$ [Figs.~3(a) and (b)]. We obtain the ground-state 
energy $e_{g} = -0.86856650(4) J$, velocity $v = 2.564(2) J$, central 
charge $c \! = \! 6 \beta/ \pi v \! = \! 1.0006(8)$, Luttinger parameter 
$K \! = \! v/4 \alpha \! = \! 1.321(1)$, and critical exponent $\nu \! = 
\! 1/(2 \! - \! K) \! = \! 1.472(2)$. This last agrees exactly with our 
gap data in Fig.~2(b), confirming the consistency and accuracy of our 
calculations. Our computed central charge is extraordinarily close to the 
expected value $c = 1$ \cite{rchs,rdeor}. Even data at the extreme precision 
we attain cannot determine whether the second derivative of $e_g$ has a 
discontinuity [Fig.~3(c)], but set a very low upper bound. A continuous 
function with a point of inflection at $D_c$ is consistent with the CFT 
expectation \cite{rthcy} that the Gaussian transition be third-order for 
$J_z = 1$.

\begin{figure}[t]
\includegraphics[width=8.5cm]{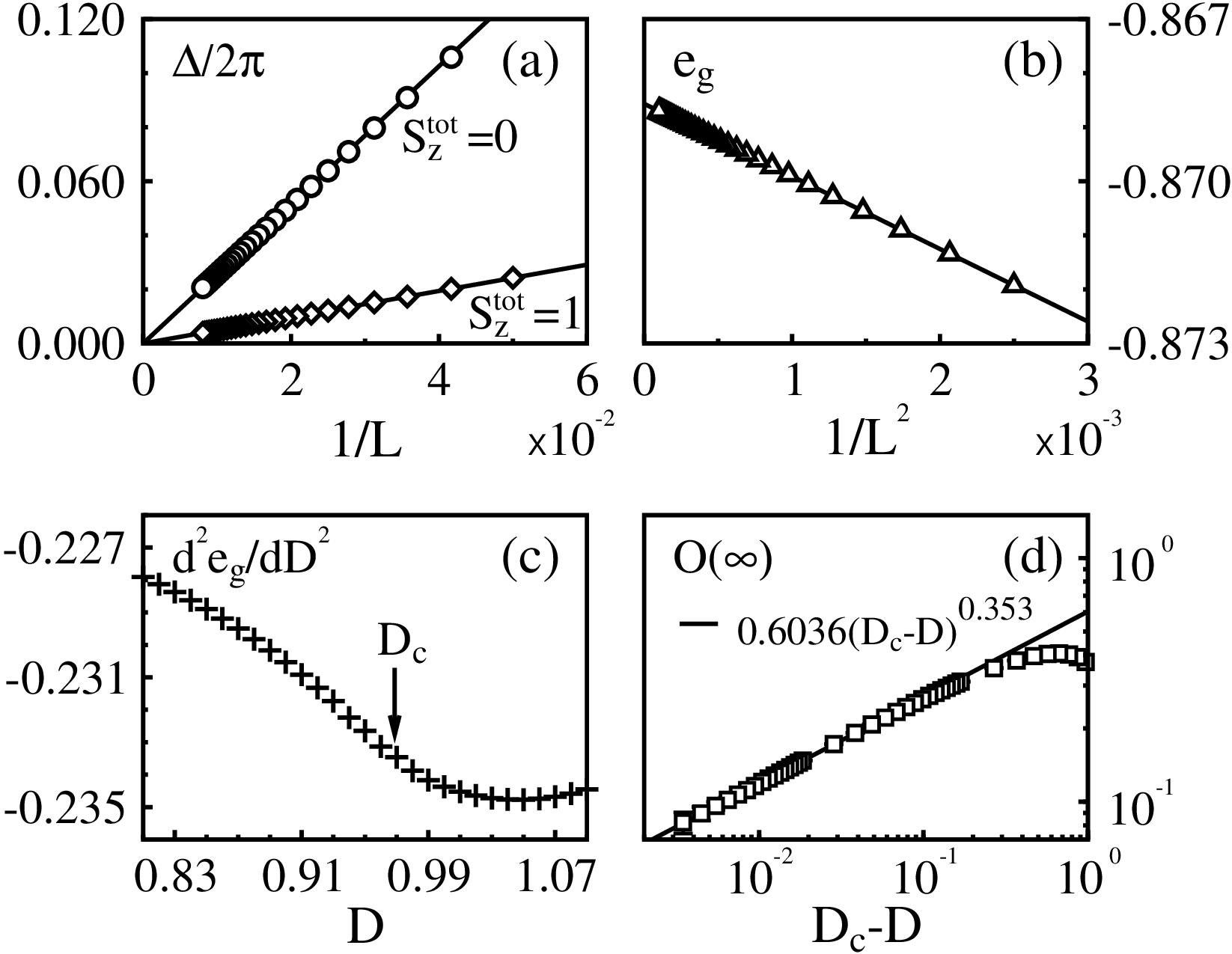}
\caption{(color online) (a) Finite-size extrapolation of lowest two gaps at 
$D_c$. Fitting lines from CFT give $\Delta (S_z^{\rm tot} = 0) = 2\pi v /L$ 
with $v = 2.564(2) J$ and $\Delta (S_z^{\rm tot} = 1) = 2\pi J \alpha /L$ with 
$\alpha = 0.48516(1)$. (b) Extrapolated ground-state energy $e_{g} $ at $D_c$, 
with CFT fit $- 0.86856650(4) J - \beta J/L^2$ and $\beta = 1.3429(4)$. (c) 
Second derivative of extrapolated energy. (d) Extrapolated transverse 
string-order parameter (see text) of the Haldane phase, with fitting line 
$0.6036(4) |D - D_c|^{0.353(1)}$. Calculations for (a) and (b) performed with 
PBCs and $m = 2000$, for (c) and (d) with OBCs and $m = 1000$.} \label{fig3}
\end{figure}

The transverse string-order parameter is defined as
\begin{equation}
 O\left( l \right) = \left \langle \hat S^{x}_{0} \exp \left( i \pi
\mbox{$\sum^{l-1}_{p=1}$} {\hat S}^{x}_{p}\right) {\hat S}^{x}_{l} 
\right \rangle,
\end{equation}
and encapsulates the incomplete $Z_{2}$$\times$$Z_{2}$ symmetry of the 
Haldane phase \cite{rls}. To reduce the complexities inherent in calculating 
this quantity, we compute correlation functions only far from the system 
boundaries \cite{schollwock}, in the left-central block $[L/4-1000, L/4]$ 
of the chain. We take the $S^{\rm tot}_z = 1$ sector as the ground state. 
Figure~3(d) shows the results of our extrapolations to infinite $L$ and $m$. 
The string-order parameter clearly shows excellent scaling behavior in the 
critical regime. The scaling exponent $\nu' = 0.353(1)$ is very close to 
the value $1/\sqrt{8}$ predicted in the 2D classical model \cite{rls}, 
demonstrating the common physics of the Gaussian, or preroughening, 
transition.

We illustrate with one example the utility of our improved DMRG calculations 
for investigating the entire Gaussian transition line. The point $J_z = 0.5$ 
has been considered by several authors \cite{rchs,rdeor,rthcy,runk}. 
Our results (Fig.~4) provide the most accurate information yet available 
for this transition: $D_c/J = 0.6355(6)$. The values of $L$ required to 
approach criticality are very much larger than for $J_z = 1$ [Fig.~4(a)], 
and the accuracy is lower because ${\cal S}(D)$ is a significantly 
flatter function [Fig.~4(b)]. Our calculations with PBCs give 
$e_g = -0.91510889(1) J$, $v = 2.185(2) J$, $c = 1.000(1)$, $K = 1.581(1)$, 
and $\nu = 2.387(5)$ at $D_c$, allowing a complete characterization of the 
physics of continuously varying exponents. 

We have considered the entropy ${\cal S} (m,L)$ at finite $m$ and $L$. 
In fact our results in Fig.~1 for $m = 1000$ and $L = 10000$ are fully 
converged for all values of $D$ outside the very narrow region $0.94 < 
D < 1.00$. We can deduce the critical behavior of ${\cal S}$ around $D_c$ 
from a massive quantum field theory \cite{rcc}, in which ${\cal S} =  
(c/6) \ln \xi + A$ with $\xi = v/\Delta$ the correlation length and 
$\Delta \propto |D - D_c|^\nu$. The convergent behavior of our data near 
$D_c$ gives exactly the critical form ${\cal S} = {\cal S}_0 - (c \nu/6) 
\ln |D - D_c|$, which is shown as the solid lines diverging at $D_c$ 
in Figs.~1 and 4. 

\begin{figure}[t]
\includegraphics[width=8.5cm]{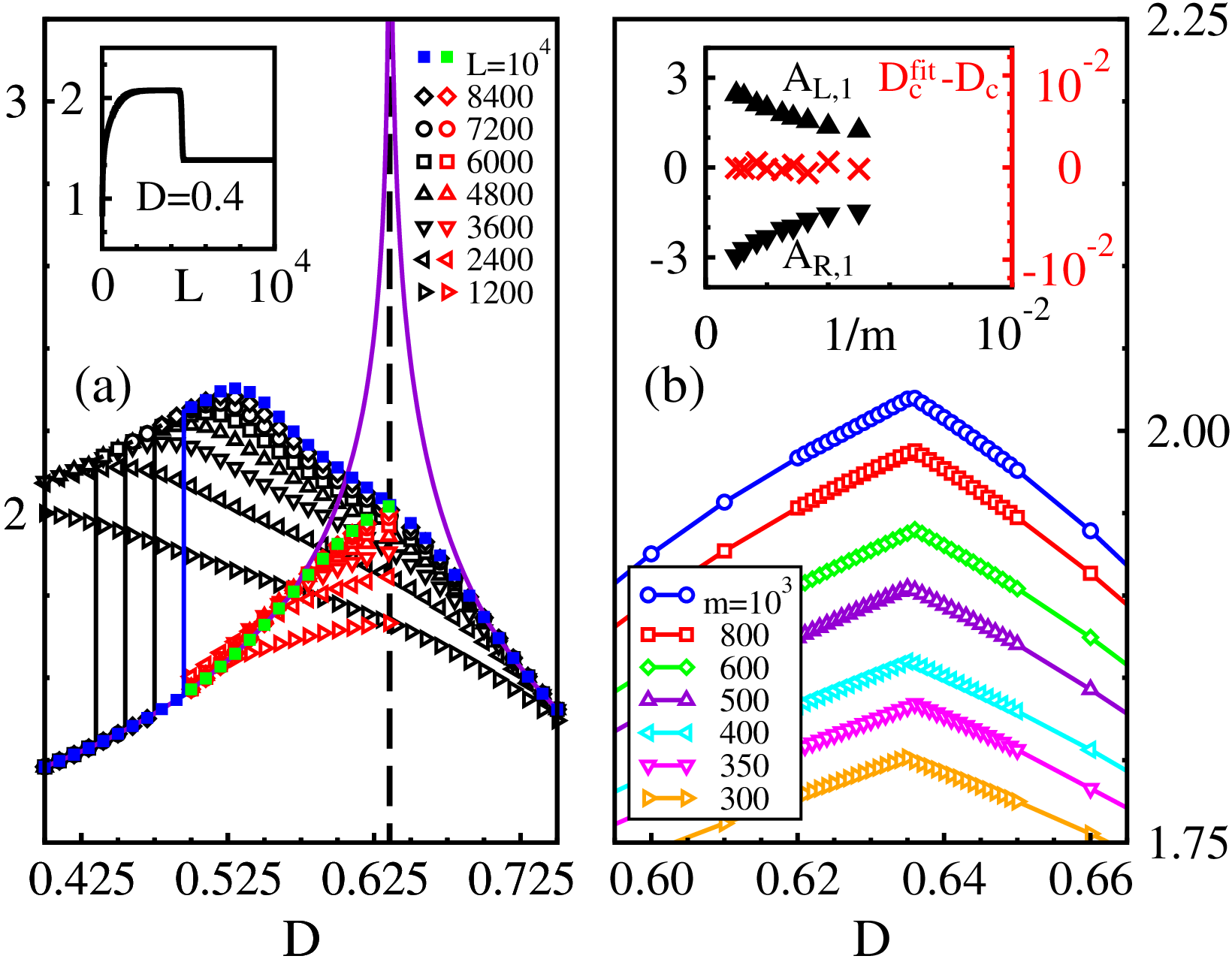}
\caption{(color online) $S(D)$ as in Fig.~1 for $J_z = 0.5$. (a) Values of 
$L$ as indicated. Inset: $\ln 2$ drop in ${\cal S}(L)$ for $D = 0.4$. (b) 
Values of $m$ as indicated.}
\label{fig4}
\vspace{-0.3cm}
\end{figure}

The Gaussian transition in the $S = 1$ chain is topological, in that the 
parity of the ground state changes from negative in the Haldane phase to 
positive in the large-$D$ phase. The transition is thus associated with a 
change in the topological spin Berry phase from $\pi$ to 0 \cite{rtajn}, 
and can be followed by a method of crossing energy levels (of states in 
the appropriate parity sectors). Our high-precision results demonstrate 
that this is indeed a very sensitive indicator of a topological transition: 
among all previous studies \cite{rbjk,rst,rgjl,rtnk,rchs,rdeor,rthcy,runk,
raho}, we find that the only accurate estimate of $D_c$ was obtained, despite 
being limited to 16-site systems, by employing this approach~\cite{rchs}.

We have demonstrated that the entropy is very valuable for discussing 
continuous phase transitions between gapped states. Many other types of 
strongly interacting quantum system fall in this category, one good
example with electronic degrees of freedom being the ionic Hubbard 
model (IHM) \cite{rmmns}. The numerically challenging transition in 
this case is of KT type. Continuous gapped-to-gapped transitions for both 
bosonic and fermionic systems exist in ultra-cold atomic condensates on 
optical lattices. The Gaussian transition has not yet been observed in 
experiment, due to difficulties in controlling the ratio $D/J$ in condensed 
matter systems, and cold-atom experiments may offer a clean solution to this 
problem.

To summarize, we calculate the critical point of the spin-one Heisenberg 
chain with single-ion anisotropy, $D_c/J = 0.96845(8)$, to extremely high
accuracy. To achieve this we introduce an improved DMRG scheme, which controls 
the absolute error of a large system and allows the elimination of end-spin 
effects. We exploit this accuracy to deduce the critical properties of many 
quantities at the Gaussian transition. The energy, entropy, and gap all 
show good scaling behavior with a single critical exponent, $\nu = 1.472(2)$. 
We apply our technique also at $J_z = 0.5$ to illustrate the continuous 
variability of exponents on the Gaussian transition line.

We thank A.~A.~Aligia and A.~Honecker for helpful discussions. This work was 
supported by the National Science Foundation of China under Grant No.~10874244 
and by Chinese National Basic Research Project No.~2007CB925001.

\end{document}